\documentclass[aps,prx,reprint,superscriptaddress,floatfix,longbibliography]{revtex4-1}
\usepackage{lipsum,amsmath,amssymb}
\usepackage{graphicx,color}

\makeatletter
\let\cat@comma@active\@empty
\makeatother
\begin{document}


\title{The impact of contact inhibition on collective cell migration and proliferation}


\author{H.P. Jain}
\affiliation{Institute of Scientific Computing, Technische Universit\"at Dresden, 01062 Dresden, Germany}
\author{D. Wenzel}
\affiliation{Institute of Scientific Computing, Technische Universit\"at Dresden, 01062 Dresden, Germany}
\author{A. Voigt}
\affiliation{Institute of Scientific Computing, Technische Universit\"at Dresden, 01062 Dresden, Germany}
\affiliation{Center for Systems Biology Dresden (CSBD), Pfotenhauerstr. 108, 01307 Dresden, Germany} 
\affiliation{Cluster of Excellence - Physics of Life, TU Dresden, 01062 Dresden, Germany}

\begin{abstract}
Contact inhibition limits migration and proliferation of cells in cell colonies. We consider a multiphase field model to investigate the growth dynamics of a cell colony, composed of proliferating cells. The model takes into account the mechanisms of contact inhibition of locomotion and proliferation by local mechanical interactions. We compare non-migrating and migrating cells, in order to provide a quantitative characterization of the dynamics and analyse the velocity of the colony boundary for both cases. Additionally, we measure single cell velocities, number of neighbour distributions, as well as the influence of stress and age on positions of the cells and with respect to each other. We further compare the findings with experimental data for Madin-Darby canine kidney cells.
\end{abstract}

\pacs{?}

\maketitle



Collective cell migration and proliferation play fundamental roles in embryonic development, tissue regeneration, wound healing and many disease processes. Identifying the principles behind these processes requires a multiscale approach, linking the properties of individual cells and cell-cell interactions to the emerging collective behaviour. Various modeling approaches have been considered to address this task, see \cite{Hakim_RPP_2017,Alert_ARCMP_2020,Moure_ACME_2020} for reviews. We here use a multiphase field model, see \cite{Nonomura_PLOS_2012,Camley_PNAS_2014,Palmieri_SR_2015,Mueller_PRL_2019,Wenzel_JCP_2019,Loewe_PRL_2020,Wenzel_arXiv_2021}, which allows for cell deformations and detailed cell-cell interactions, as well as subcellular details to resolve the mechanochemical interactions underlying cell migration. Also topological changes, such as T1 transitions, follow naturally in this framework. Multiphase field models, together with efficient numerics and appropriate computing power, allow to analyse the connection of single cell behaviour with collective migration and growth of cell colonies, at least for moderate numbers of cells. They already led to quantitative predictions of many generic features in multicellular systems \cite{Peyret_BJ_2019,balasubramaniamEtAl2021,Wenzel_arXiv_2021}. 

\begin{figure}[htb]
    \centering
    \includegraphics[width=0.238\textwidth]{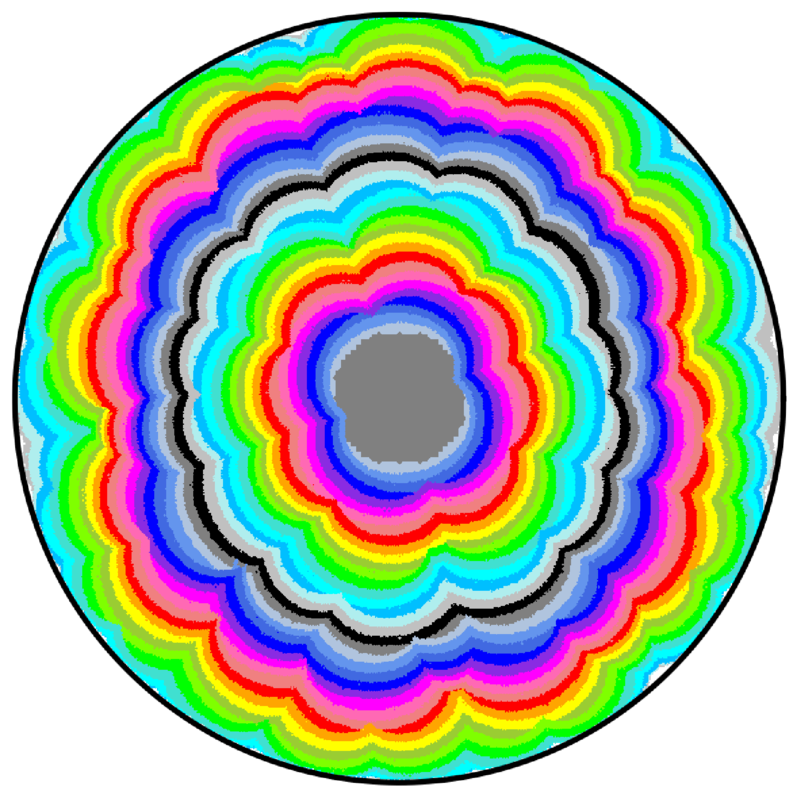} 
    \includegraphics[width=0.238\textwidth]{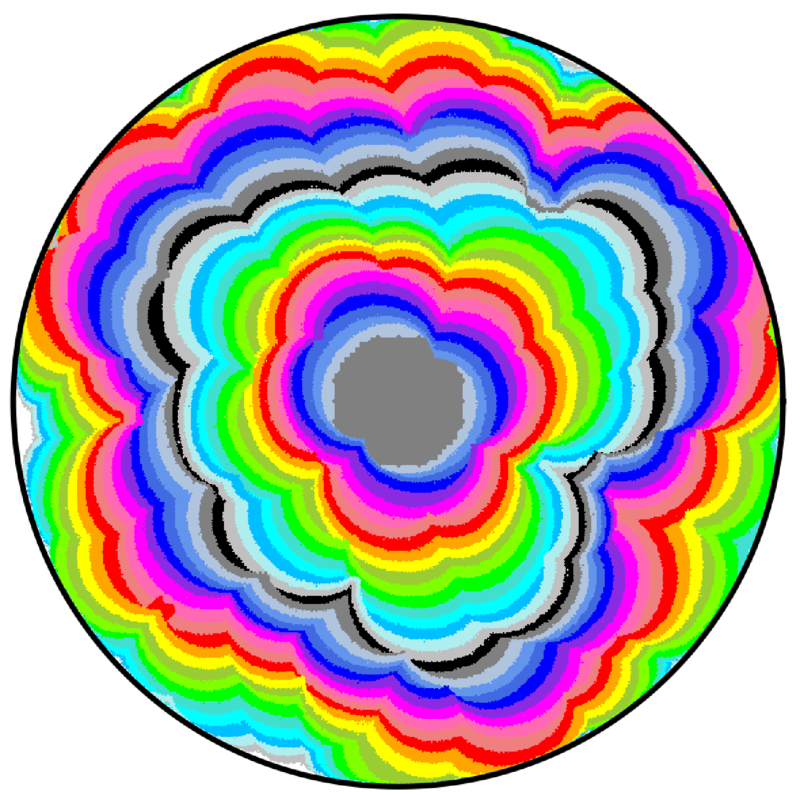} 
    \caption{Superimposed snapshots of cell colony at different times, starting with one circular cell in the centre (on top, gray) at $t = 1$. (left) non-migrating, (right) migrating cells. The outer circle marks the confinement. For selected snapshots, also showing the individual cells of the colony, see SI.}
    \label{fig1}
\end{figure}

Here, we especially focus on the complex interaction of migration and proliferation, which are regulated by contact inhibition processes. We distinguish between two different inhibitory mechanisms. On the one hand, we consider contact inhibition by locomotion (CIL), which describes the tendency of cells to stop migration, deform and change direction when coming in contact with other cells, see \cite{Stramer_NRMCB_2017} for a review. On the other hand, we also investigate contact inhibition of proliferation (CIP), which refers to the suppression of cell growth and divisions in dense regions of tissues \cite{Fisher_Science_1967,Stoker_Nature_1967}. CIL naturally results from mechanical interactions of the multiphase field model and its impact on collective migration is well explored, within this framework \cite{Marth_IF_2016,Mueller_PRL_2019,Wenzel_JCP_2019,Peyret_BJ_2019,Wenzel_arXiv_2021}. In contrast to CIL, proliferation cannot easily be incorporated in the energetic description of a multiphase field model. Even if such attempts on a single cell level exist \cite{Zwicker_NP_2017,Li_JSM_2020}, we here follow an adhoc procedure, which simply divides a cell if it has reached a certain size \cite{Drasdo_PRE_1995,Nonomura_PLOS_2012,Basan_PNAS_2013,Aland_BPJ_2015}. CIP is included by a growth factor for each cell, which is affected by cell-cell interactions. We study the interplay of CIL and CIP for the dynamics of small groups of cells, proliferating in a circular confinement, similar to the experiments on monolayers of epithelia cells in \cite{Puliafito_PNAS_2012,Doxzen_IB_2013}. In particular, we analyse emerging features, such as the speed of colony growth, the velocity of individual cells, topological measures, such as number of neighbor distributions, as well as relations between stress and age on position of the cells in the colony. For other computational studies on CIL and CIP, but with much stronger restrictions on the deformation of cells, we refer to \cite{Puliafito_PNAS_2012,Aland_BPJ_2015,George_SR_2017,Schnyder_SR_2020}. Fig. \ref{fig1} shows the dynamics of a growing colony for non-migrating and migrating cells, already indicating the differences of the models with respect to morphology and velocity of the colony boundary.

We consider two-dimensional phase field variables $\phi_i(\mathbf{x},t)$, one for each cell. Values of $\phi_i = 1$ and $\phi_i = -1$ denote the interior and the exterior of a cell, respectively. The cell boundary is defined implicitly by the region $-1 < \phi_i <  1$. The dynamics for each $\phi_i$ is considered as
\begin{equation}
    \partial_t \phi_i + v_0 (\mathbf{v}_i \cdot \nabla \phi_i) = \Delta \frac{\delta \mathcal{F} }{\delta \phi_i} + k_i (\phi_i + 1), 
    \label{eq:phi}
\end{equation}
for $i = 1, \ldots, N$, where $N = N(t)$ denotes the number of active cells. $\mathcal{F}$ is a free energy and $\mathbf{v}_i$ a vector field used to incorporate active components, with a self-propulsion strength $v_0$. For non-migrating cells we set $v_0 = 0$. All quantities are non-dimensional quantities. We consider conserved dynamics and add an exponential growth term with a growth rate $k_i$. The free energy $\mathcal{F} = \mathcal{F}_{CH} + \mathcal{F}_{INT} + \mathcal{F}_{CON}$ contains passive contributions, where
\begin{eqnarray}
\mathcal{F}_{CH} &=& \sum_{i=1}^N \frac{1}{Ca}\int_\Omega \frac{\epsilon}{2}\|\nabla\phi_i\|^2 + \frac{1}{\epsilon}W(\phi_i)\,\text{d}\mathbf{x}, \\
\label{eq:IntEnergy}
\mathcal{F}_{INT} &=& \sum_{i=1}^N \frac{1}{In}\int_\Omega B(\phi_i) \sum_{j\neq i} w(\phi_j)\,\text{d}\mathbf{x}, \\
\mathcal{F}_{CON} &=& \sum_{i=1}^N \frac{1}{Co}\int_\Omega B(\phi_i) w(\phi_{con})\,\text{d}\mathbf{x},
\end{eqnarray}
with non-dimensional capillary, interaction and confinement number, $Ca$, $In$ and $Co$, respectively. The first is a Cahn-Hilliard energy, with $W(\phi_i) = \frac{1}{4}(\phi_i^2 - 1)^2$ a double-well potential and $\epsilon$ a small parameter determining the width of the diffuse interface. This energy stabilizes the cell interface. For simplicity, we here neglect other properties of the cell boundary, e.g., bending forces. In \cite{Marth_JRSI_2015} they are shown to be negligible in the context of cell migration. The second is an interaction energy with $B(\phi_i) = \frac{\phi_i + 1}{2}$, a simple shift of $\phi_i$ to values in $[0,1]$ and a cell-cell interaction potential 
\begin{equation}
    w(\phi_j) = 1 - (a+1)\left(\frac{\phi_j -1}{2}\right)^2 + a \left(\frac{\phi_j -1}{2}\right)^4
    \label{eq:interaction}
\end{equation}
approximating a short range potential, which is only active within the interior of the cell and its diffuse boundary. The approach offers the possibility to consider repulsive as well as attractive interactions which can be tuned by parameter $a$, see SI. The last energy models the interaction with the confinement, which is given by the phase field function 
$\phi_{con}(\mathbf{x}) = \tanh ((\| \mathbf{x} - \mathbf{m}\| - r_{con} ) / (\sqrt{2} \epsilon))$, with $\mathbf{m} = (l/2,l/2)$ the centre of the computational domain and $r_{con}$ the radius of the circular confinement. The interaction potential is the same as in eq. \eqref{eq:interaction}, but we only consider repulsive interactions, see SI. Both $\mathcal{F}_{INT}$ and $\mathcal{F}_{CON}$ provide the mechanical source for CIL.  

For the definition of $\mathbf{v}_i$ in eq. (\ref{eq:phi}), we follow the simplest possible approach, which can be considered as a generalization of active Browning particles \cite{Fily_PRL_2012,Redner_PRE_2013,Wysocki_EPL_2014} to deformable objects \cite{Loewe_PRL_2020}. In this approach the propulsion speed is the same for each cell, but the direction of motion, determined by the angle $\theta_i$ is controlled by rotational noise $d \theta_i(t) = \sqrt{2 D_r} d W_i(t)$, with diffusivity $D_r$ and a Wiener process $W_i$, which results in $\mathbf{v}_i = (\cos{\theta_i}, \sin{\theta_i})$. The growth rate $k_i$ is sampled from a Gaussian distribution with mean and variance determined by the product of a constant growth factor $k_{growth}$ and CIP factors for cell-cell interaction $f_i \in [0,1]$ and cell-confinement interaction $\xi_i \in [0,1]$. As a measure for contact, we consider the variational derivatives $\delta \mathcal{F}_{INT} / \delta \phi_i$ and $\delta \mathcal{F}_{CON} / \delta \phi_i$ and define the total interactions, $T_i = \int_\Omega \delta \mathcal{F}_{INT} / \delta \phi_i \text{d}\mathbf{x}$ and $T_{con,i} = \int_\Omega \delta \mathcal{F}_{CON} / \delta \phi_i \text{d}\mathbf{x}$, which enter in 
\begin{align}
    f_i &= \max (0, \min (1 - \text{sign} (T_i) \left(\frac{T_i}{L}\right)^2, 1))
\end{align}
and $\xi_i$ defined in the same way, with $T_i$ replaced by $T_{con,i}$. The parameter $L$ is a limiting factor, see SI.

Cell division is manually introduced if the cell size reaches a threshold. Following experimental evidence for epithelia tissue \cite{Wyatt_PNAS_2015}, we divide the cell perpendicular to its elongation axis. The number $N$ of active cells is increased by one and the age of the two daughter cells start from zero. The overall size of the two daughter cells is slightly below that of the mother cell due to the necessity to introduce a diffuse interface between the two daughter cells. For further details, we refer to SI.


We employ a parallel and adaptive finite element method to solve the $N$ coupled systems of partial differential equations for $\phi_i$. The algorithm is implemented in AMDiS \cite{Vey_CVS_2007,Witkowski_ACM_2015} and the algorithmic concepts to achieve parallel scaling with the number of cells $N$ are described in \cite{Praetorius_NIC_2017}. Briefly, they consider one core for the evolution of each cell and parallel concepts from particle methods to reduce the communication overhead due to cell-cell interaction. The coupled fourth order equations in eq. \eqref{eq:phi} are split into coupled systems of second order equations for the phase field variables $\phi_i$ and the corresponding chemical potentials $\mu_i = \delta \mathcal{F} / \delta \phi_i$. The discretization in time is semi-implicit with a Taylor expansion of the double-well potential and an explicit treatment of the other non-linear terms. For the Cahn-Hilliard part we use an additional stabilisation \cite{Raetz_JCP_2006,Backofen_IJNAM_2019,Salvalaglio_MMAS_2021} to ensure $\phi_i \in [-1,1]$ and to increase accuracy.

\begin{figure}[htb]
    \centering
    \includegraphics[width=0.4\textwidth]{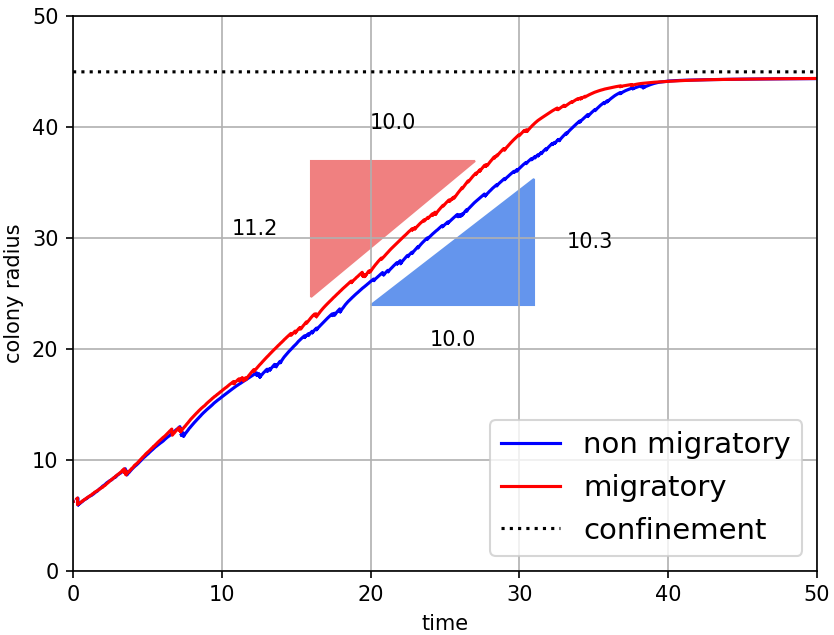}
    \caption{Radius of a circle of corresponding area of growing colony over time and corresponding slope for $In = 0.05$ and $L = 10,000$. The horizontal line shows the confinement, which is reached at $t \approx 40$. For related results with varying $In$ and $L$ see SI.}
    \label{fig2}
\end{figure}

\begin{figure}[htb]
    \centering
    \includegraphics[width=0.47\textwidth]{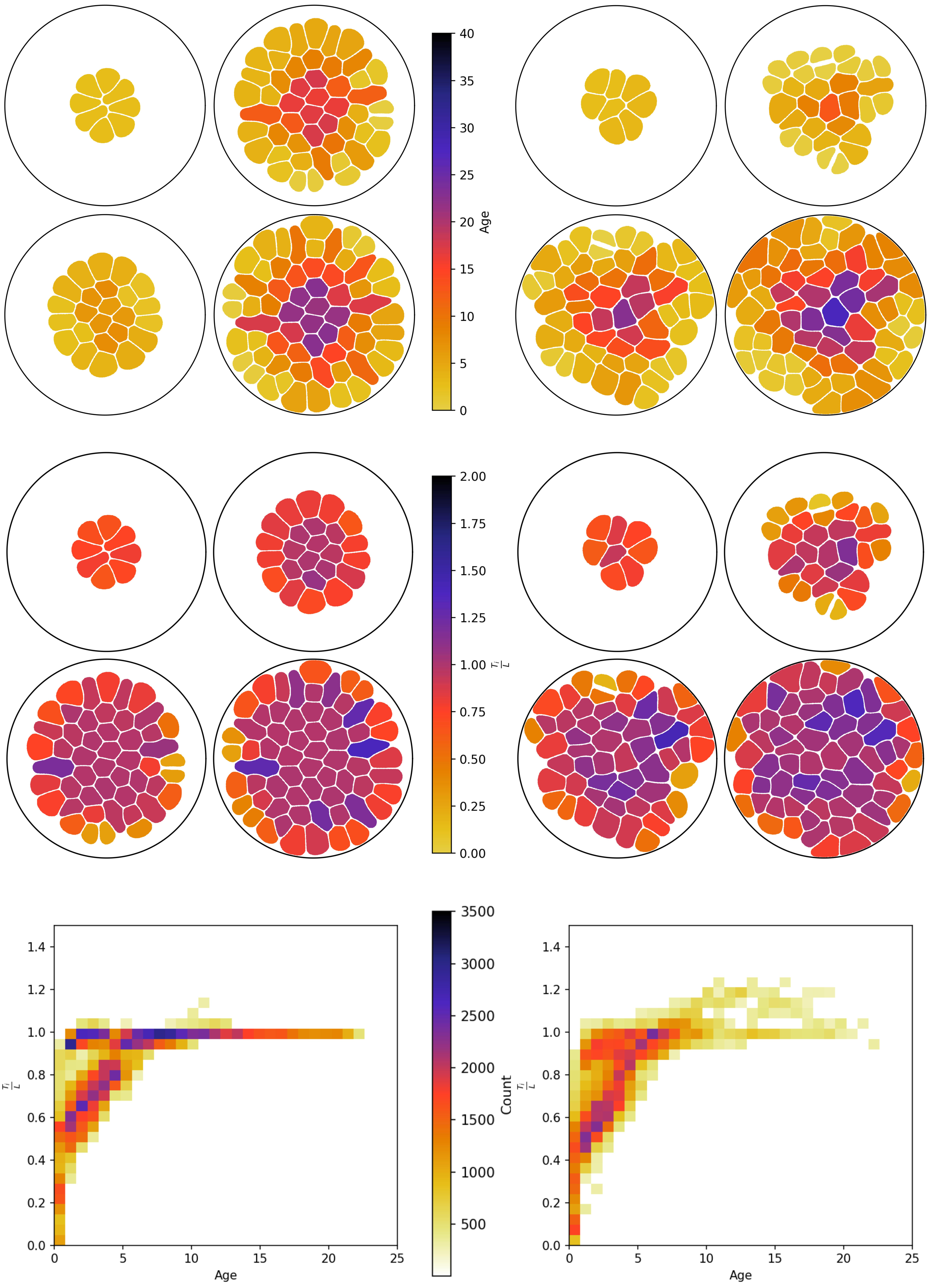}
    \caption{Four snapshots of cell colony growth with color coding by age (top) and stress $T_i / L$ (middle) for non-migrating cells (left) and migratiing cells (right). The corresponding times are $t = \{10,20,30,35\}$. (bottom) Corresponding relation between age and stress for all cells at all times $t \leq 35$. The color coding is with respect to the number of cells with the considered age/stress relation. Only points with more than 250 counts are shown.}
    \label{fig3}
\end{figure}

We start with one circular cell, $N = 1$, in the centre of a circular confinement within the computational domain, $\Omega = [0,l] \times [0,l]$. The size of the cell is below the dividing threshold. Numerical parameters, such as grid resolution and time step are considered as large as possible to ensure stable behaviour and resolution of the essential physics. The grid spacing within the diffuse boundary is $h \approx 0.2 \epsilon$, in the interior of each cell $h \leq \epsilon$ and in the exterior $h \leq 10\epsilon$ with increasing values for regions far away from the interior. The time step is chosen as $\tau = 0.005$. Other parameters, if not specified differently, are set as $l = 100$, $r_{con} = 45$, $\epsilon = 0.15$, $Ca = 0.1$, $Co = 0.005$, $D_r=0.01$ and $k_{growth} = 0.3$. We compare non-migrating and migrating cells ($v_0 = \{0, 1\}$) and consider $In = 0.05$ and $L = 10,000$. Fig. \ref{fig2} shows the growth, expressed through the radius of a circle with the same area as the cell colony. The radius grows linearly, and slightly faster for the migrating cells. This linear growth regime corresponds to a constant boundary velocity of the cell colony and is consistent with experimental results for confluent MDCK cells \cite{Puliafito_PNAS_2012} and theoretical approaches, in which, due to CIP, only the cells at the boundary are able to grow and divide \cite{Schnyder_SR_2020}. In order to further analyse the effect of CIL and CIP on colony growth, we consider the age and the stress of each cell, see Fig. \ref{fig3}. Stress is measured in terms of the total interaction $T_i$, normalized by the limiting factor $L$. While for the non-migrating case the age of the cells is more or less decreasing with increasing radius of the cell colony and the inner cells develop a homogeneous stress distribution, both age and stress are more heterogeneously distributed in the migrating case, which is quantified in the age/stress diagrams. For cells with age below five, both diagrams are similar, indicating an increase of stress with age. But for cells with age above five, the diagrams differ. The homogeneous stress distribution for non-migrating cells results in a horizontal line at $T_i/L \approx 1$. For migrating cells, this is less pronounced with larger stresses emerging, corresponding to a heterogeneous distribution. Due to migration cells can generate space, which allow them to grow further if the threshold for cell division is not reached. This increases the cell density and leads to stresses with $T_i/L > 1$.

\begin{figure}[htb]
    \centering
    \includegraphics[width=0.47\textwidth]{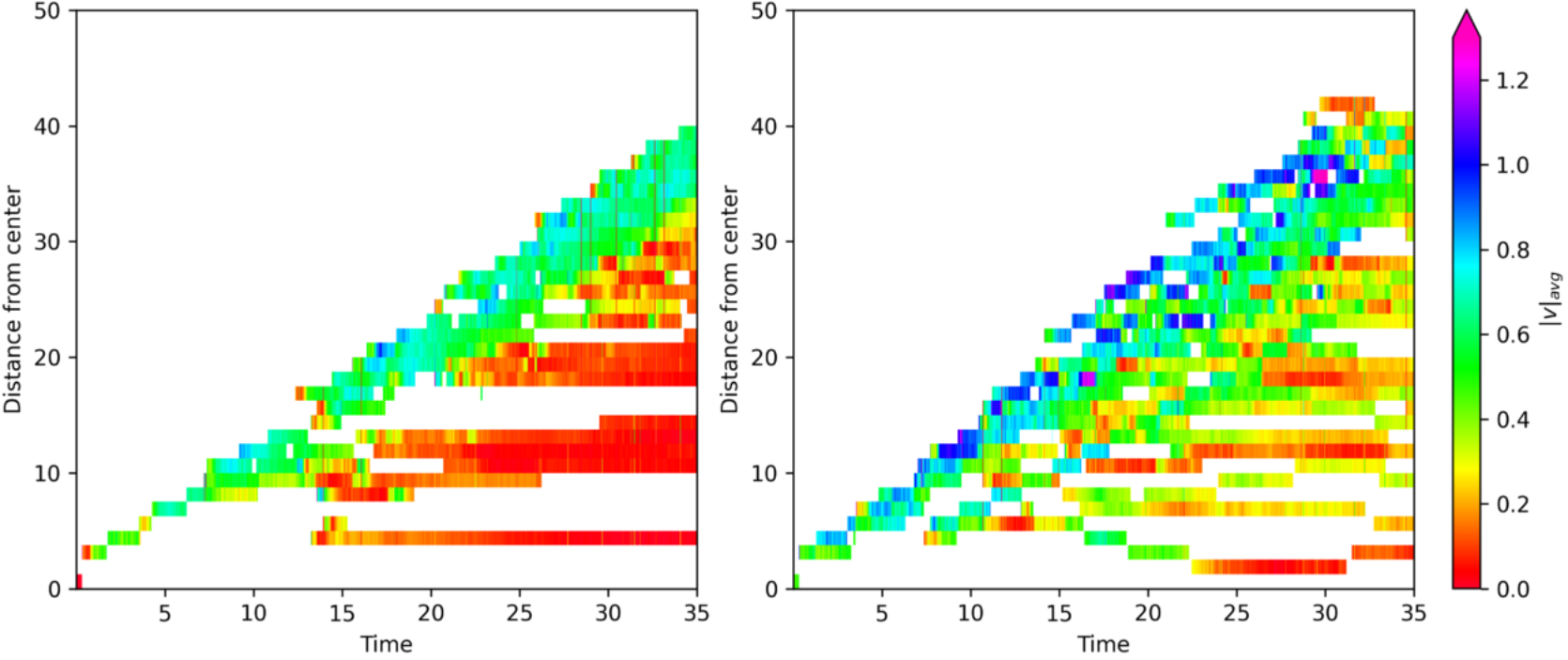}
    \includegraphics[width=0.47\textwidth]{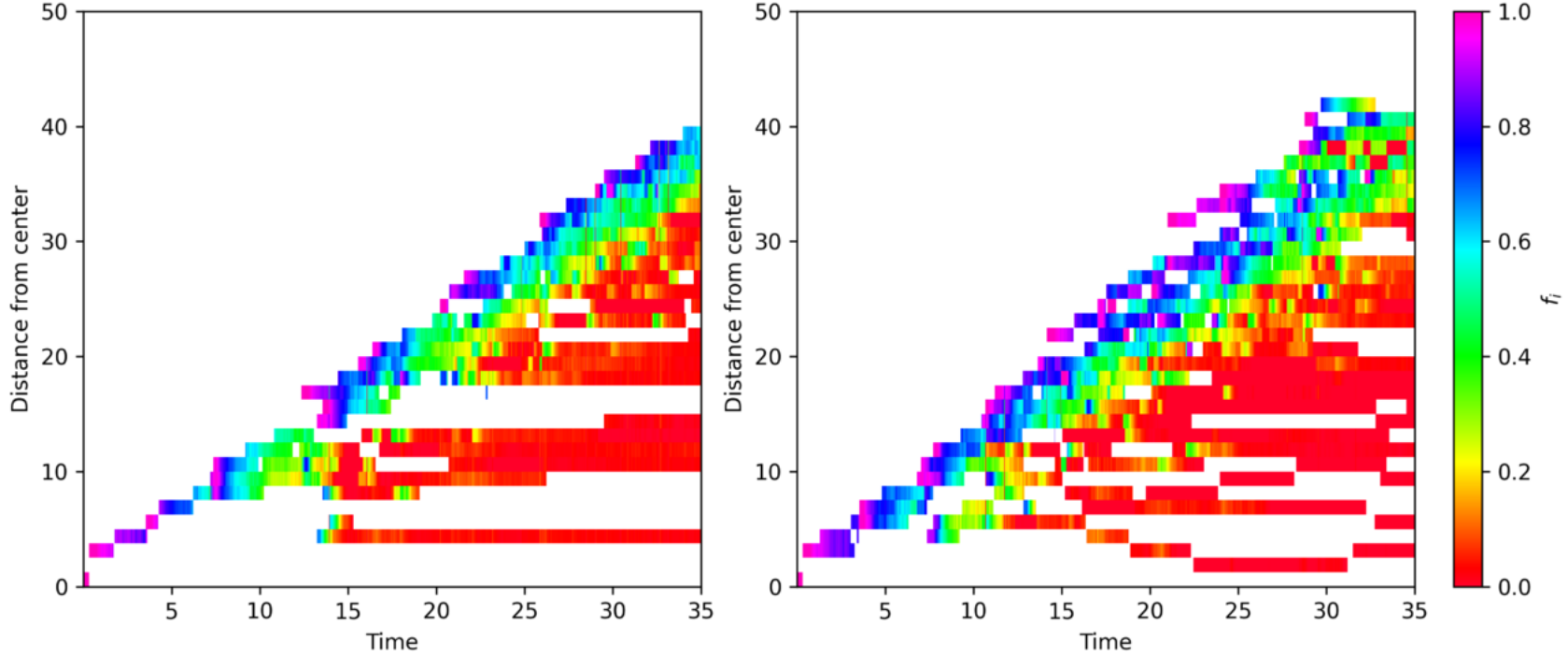}
    
    \caption{Radial analysis. (top) Magnitude of cell velocity $|\mathbf{v}|_{avg}$ as a function of distance from colony centre and time. (bottom) Measure of cell-cell interaction $f_i$. (left) non-migrating, (right) migrating cells.}
    \label{fig4}
\end{figure}

The increased heterogeneity for the migrating case also becomes evident if the cell properties are averaged according to their position in the colony with respect to the center. Fig. \ref{fig4} shows the magnitude of the cell velocity and the measure for cell-cell interaction $f_i$. While spatial-temporal distribution of the cell-cell interaction is similar for the non-migrating and migrating cases, the cell velocities strongly differ. Not only within the centre of the colony, where the non-migrating cells are more or less stationary and significant movement is present for the migrating cells, but also on the colony boundary, which shows significantly larger velocities for the migrating cells. 
\begin{figure*}[htb]
    \centering
    \includegraphics[width=0.32\textwidth]{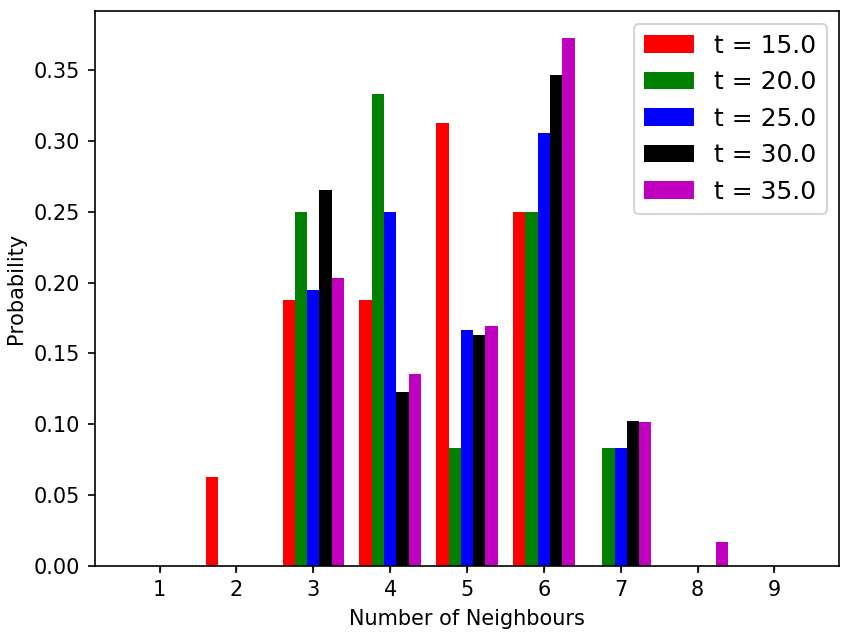}
    \includegraphics[width=0.32\textwidth]{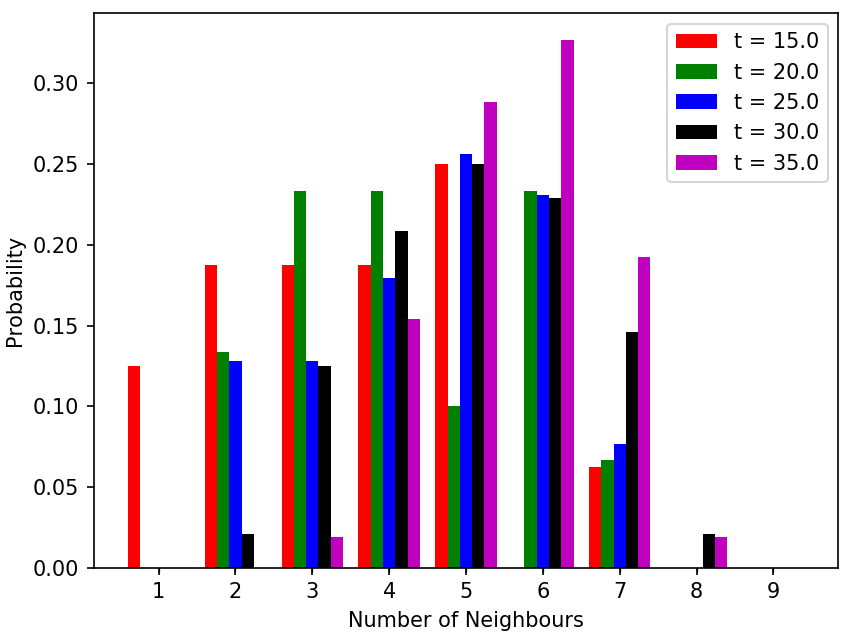}
    \includegraphics[width=0.32\textwidth]{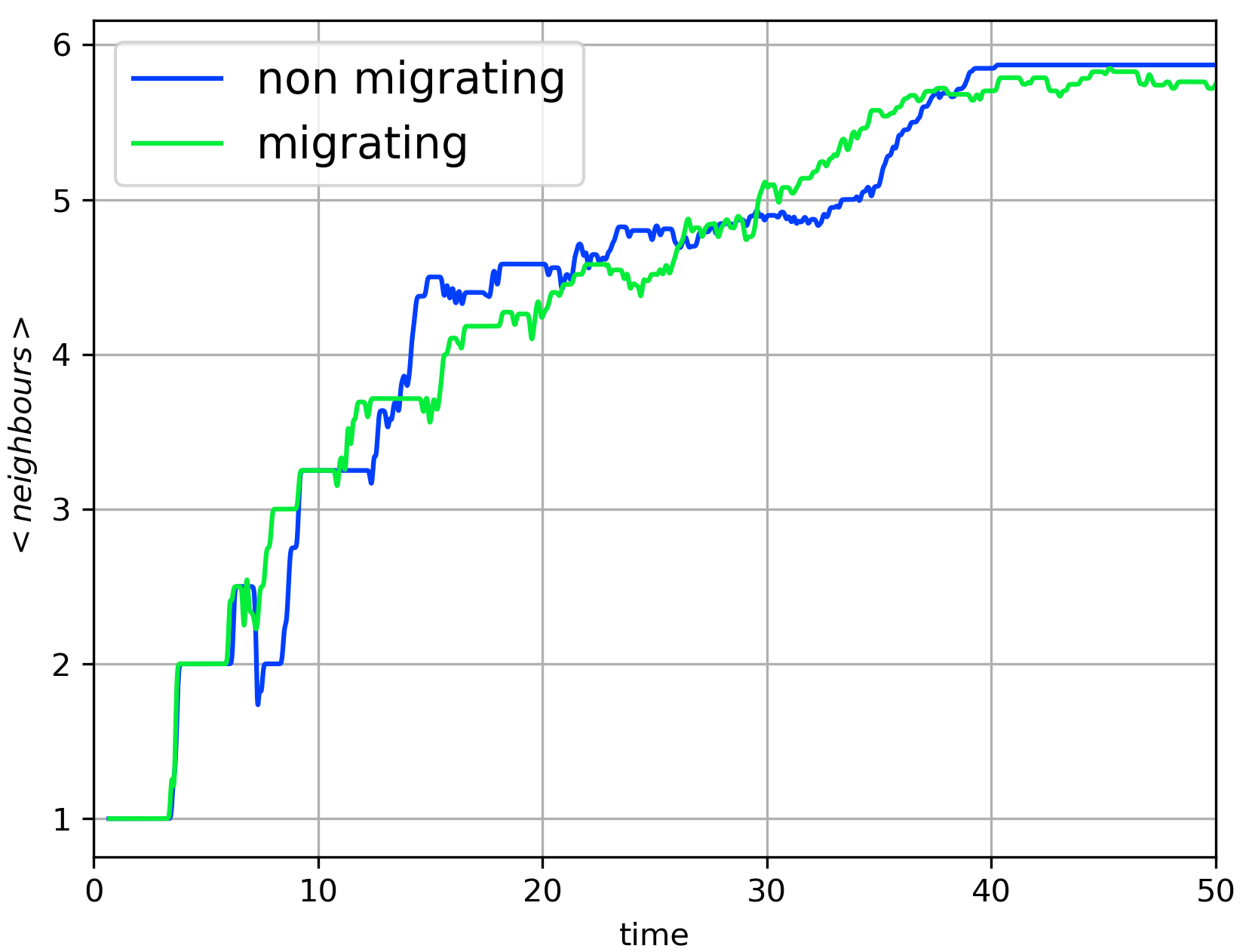}
    \caption{Histogram of the coordination number of the cells at different times: (left) non-migrating, (middle) migrating cells. (right) Averaged number of neighbors over time for both cases.}
    \label{fig5}
\end{figure*}
According to \cite{Schnyder_SR_2020} this larger velocity at the colony boundary is the main reason for the increased growth rate in the migrating case. It is argued that at long times the colony expands as fast as the cells on the boundary are able to migrate away from the colony centre. 

Another measure concerns the topology of the cells in the colony. We consider the coordination number, the number of nearest neighbours, for each cell and calculate the histogram, see Fig. \ref{fig5}. The data is shown for different time instances. As time progresses, cells with six neighbours become the most prevalent. However, also the histogram shows the difference between the non-migrating and migrating cases. While for non-migrating cells an equilibrium-like homogeneous configuration with cells with mainly six neighbors emerges, the structure of the migrating cells is more heterogeneous with significantly more cells with five and seven neighbors. This again is in agreement with measurements for MDCK cells \cite{Puliafito_PNAS_2012}. In contrast to the simplified models in \cite{Schnyder_SR_2020}, which overestimate configurations with six neighbours at long times, our simulations lead to more realistic distributions. This might be due to the larger flexibility in cell deformation. The reached distribution for migrating cells is also in good agreement with the mathematically predicted equilibrium distribution of cellular polygons, which is empirically confirmed in tissue samples from vertebrate, arthropod and cnidarian organisms \cite{Gibson_Nature_2006}. Detailed validations on neighbor distributions in related multiphase field models without growth and division have already been shown in \cite{Wenzel_JCP_2019}. On average, the number of neighbors evolves similarly for the non-migrating and the migrating cells, see Fig. \ref{fig5}(right), and reaches six, which is consistent with Euler's polyhedral formula. This shows that independent of migration, a predominantly hexagonal topology emerges as a consequence of cell division, a phenomena also observed in plant tissue \cite{Lewis_AR_1926}.

In summary, we have used a multiphase field model to explore the effect of CIL and CIP in growing colonies of non-migrating and migrating cells. The extension of the multiphase field model, as a minimal cell-based model, which accounts for cell deformability and force transmission at cell-cell contacts, to growth and cell division, allows to analyse the impact of contact inhibition. While CIL is naturally considered in the multiphase field model, CIP is included by linking cellular growth to the short-range interaction with neighboring cells, which is realised by considering the chemical potential associated with the interaction energies. The stochastic behaviour of migration and growth on a single cell level manifests in global patterns in a multicellular context. The emerging neighbor distribution in the growing colony is a result of cell division and in good agreement with experimental measurements for various organisms. The proposed model leads to the typical linear boundary growth of the colony radius and allows for various investigations concerning cell velocity, age and stress distributions. They all significantly differ between non-migrating and migrating cells and ask for experimental validations.\\


A.V. acknowledges support by the German Research Foundation (DFG) under Grant FOR3013. We further acknowledge computing resources provided at J\"ulich Supercomputing Center under Grant No. PFAMDIS. \\

\bibliography{library}

\end{document}